\documentclass[a4paper,conference]{IEEEtran} 
\ifCLASSINFOpdf
\else
\fi
\usepackage[cmex10]{amsmath}
\usepackage{amssymb,verbatim}
\usepackage{graphicx}

\begin{document}
%
\title{%
Towards Quantum Enigma Cipher II\\
-A  protocol  based on quantum illumination-
}

\author{
\IEEEauthorblockN{Osamu Hirota\\}
\IEEEauthorblockA{
Quantum ICT Research Institute, Tamagawa University\\
6-1-1 Tamagawa-gakuen, Machida, Tokyo 194-8610, Japan\\
{\footnotesize\tt E-mail: hirota@lab.tamagawa.ac.jp} \vspace*{-2.64ex}}
}

\maketitle

\begin{abstract}
This research note II introduces a way to understand a basic concept
 of the quantum enigma cipher.  
The conventional cipher is designed by a mathematical algorithm and 
its security is evaluated by the complexity of the algorithm in  
 security analysis and ability of computers. 
This kind of cipher can be decrypted with probability one in principle by 
the Brute force attack in which an eavesdropper tries all the possible keys 
based on the correct ciphertext and some known plaintext.
A cipher with quantum effects in physical layer may protect the system from 
the Brute force attack by means of the quantum no cloning theorem 
and randomizations based on quantum noise effect. The randomizations for 
the ciphertext which is the output from the mathematical encryption box
 is crucial to realize a quantum enigma cipher. 
Especially, by randomizations, it is necessary to make a substantial 
 difference in accuracy of ciphertext in eavesdropper's observation
  and legitimate user's observation. 
The quantum illumination protocol can make a difference in error performance
 of the legitimate's receiver and the eavesdropper's receiver. 
 This difference is due to differences in ability of the 
legitimate's receiver with entanglement and the eavesdropper's 
receiver without entanglement.
It is shown in this note that the quantum illumination can
 be employed as an element of the most simple quantum enigma cipher.
 
\end{abstract}

%
\IEEEpeerreviewmaketitle
\section{Introduction}
  The general network systems need to be protected from interception by 
unauthorized parties. The most serious attack is 
 `` Cyber attack against Layer-1 (physical layer such as optical 
communication line)", because  technologies of coupler for tapping have been 
developed by several institutes [1]. In addition, there are many optical 
monitor ports for network maintenance. In fact physical layer
 of high speed data link is a defenseless. 
 To date, that protection has been provided by classical encryption systems.
 However, such technologies cannot ensure the provable security,
and also the eavesdropper can obtain the correct ciphertext:$C$ of 
mathematical cipher for payload at Layer-2, and she can store 
 it in memory devices. Thus, we cannot rule out the possibility that
 the cipher may be decrypted by future technology.
 
 The best way to protect  high speed data is to physically randomize 
 signals as the ciphertext of the mathematical encryption. 
This is called physical random cipher. 
The most important feature of this physical random cipher is that 
the eavesdropper cannot get the correct ciphertext of mathematical 
encryption box, for example a stream cipher by PRNG
 (pseudo random number generator), from 
communication lines, while the legitimate user can get it based on 
a knowledge of secret key of PRNG. Thus, the ciphertext: $Y^B(C)$ 
as the signal of the legitimate user and the ciphertext: $Y^E(C)$ 
as the signal of the eavesdropper may be different as  $Y^B(C) \ne Y^E(C)$. 

Along with this concept, Quantum Enigma Cipher allows a secure high speed 
data transmission by means of the quantum noise randomization by 
a mathematical encryption box and signal modulation systems, 
or by an integration of a mathematical encryption box and a physical 
encryption box [2].
When we consider how to realize such a system, we have to take into account 
the following requirements on the encryption system in the real world:\\
\\
{\bf{Requirement of specifications:}}\\
 {{(1) Data-speed:}}1 Gbit/sec $\sim$ 100 Gbit/sec\\
 {{(2) Distance:}} 1000 Km $\sim$ 10000 Km \\
 {{(3) Encryption scheme:}} Symmetric Key Cipher \\
 {{(4) Security:}} Provable security, Secure against Brute force attack
(exhaustive search trial for secret key) by means of computer and also 
 physical devices. \\

Recently, MIT groups claimed to be capable of their 
quantum illumination scheme to provide ultra-broadband 
secured communication [3,4]. The principle is 
that the quantum illumination protocol can produce  
the difference between Alice's error performance and Eve's 
error performance. 
We can adopt such a scheme in quantum enigma cipher
 as a physical randomization technique. The author believes that this is  
  the most appropriate application of the quantum illumination. 
 In the following sections, we will discuss the reason and  method to 
realize quantum enigma cipher based on quantum illumination.

\section{Security of symmetric key cipher including one time pad}

\subsection{Model}
Let us describe a standard symmetric key encryption. 
A general symmetric key encryption $\Lambda$ can be given by
\begin{equation}
\Lambda =([P_K], Enc, Dec)
\end{equation}
where $[P_K]$ is key generation algorithm and it provides 
key sequence $K\in {\cal{K}}$ depending on the probability $P_K$,
$Enc$ is an encryption algorithm which generates  ciphertext 
$C=Enc(K,M)$ where $M$ is plaintext, $Dec$ is a decryption 
algorithm which produces plaintext $M=Dec(K,C)$.

\subsection{Security criterion}
When $\Lambda$ cannot be decrypted by means of computational resource, 
its security is evaluated by ``Guessing probability"[5,6].\\
\\
(i) Ciphertext only attack on data:
\begin{equation}
P_G(M) = \max\limits_{M \in {\cal{M}}} P(M|C)
\end{equation}
(ii) Ciphertext only attack on key:
\begin{equation}
P_G(K) = \max\limits_{K \in {\cal{K}}} P(K|C)
\end{equation}
\\
On the other hand, when some plaintext $M_k$ and ciphertext 
corresponding to them are known, it is called known plaintext attack. 
It is easy to generalize the above formula as follows:\\
\\
(iii) Known plaintext attack on data:
\begin{equation}
P_{G_k}(M) = \max\limits_{M \in {\cal{M}}} P(M|C,M_k)
\end{equation}
(iv) Known plaintext attack on key:
\begin{equation}
P_{G_k}(K) = \max\limits_{K \in {\cal{K}}} P(K|C,M_k)
\end{equation}
\\
These are sometimes called maximum ``a posteriori probability" guessing.
If one needs an average, then one can define average guessing probability 
as follows:
\begin{equation}
\bar{P_G}(M)=\sum\limits_{C\in {\cal{C}}}P(C)
\max\limits_{M \in {\cal{M}}} P(M|C)
\end{equation}
\\
\subsection{Security of ideal one time pad}
When the distribution $P_K$ is uniform, the one time pad has the perfect
 secrecy such that 
 \begin{equation}
 P_G(M) = \max\limits_{M \in {\cal{M}}} P(M|C)=P(M)
 \end{equation}
 \\
 However, even if the system has the perfect secrecy, it does not mean 
 ``secure" against known plaintext attack on data when data is a language 
 such as English. That is,\\
 \\
 $\lceil$The perfect secrecy means secure against ciphertext only attack, 
and it does not imply the security against
  ``known plaintext attack and falsification attack".$\rfloor$ \\
 \\
 Thus, the term of ``unconditional security" is misleading.
 Let us show an example. 
 The eavesdropper can get the correct chiphertext of the length $|K|$ bits, 
 and she can launch the Brute force attack. The decrypted data sequences
 of the length  $|K|$ bits give all combination of English alphabet
  (ASCII code) of length $|K|$ bits.
 These include a large number of correct English words such as
 ``orange, signal, cipher, and so on". When the attack is 
ciphertext only attack, she cannot decide 
 which word is the real plaintext. 
However, if she knows the first alphabet ``o" 
 as the known plaintext attack, the correct word may be ``orange".
Thus, the guessing probability may become very large value.

 \subsection{Security of one time pad forwarded by QKD}
 The quantum key distribution does not provide the perfectly uniform 
 distribution for key sequence $K_G$ against an eavesdropper.
 In fact, the average guessing probability is given by Portman and Renner[7]
  as follows:
 \begin{equation}
\bar{P_G}(K_G) \le \frac{1}{2^{|K_G|}} + d 
\end{equation}
where $d$ is the trace distance in QKD protocol.
Thus, the one time pad forwarded by QKD is non ideal one time pad 
which is encrypted by key sequence with non uniform distribution.
 That is, 
\begin{equation}
\Lambda =([P_K]\ne ideal, Enc, Dec), P_K \ne \frac{1}{2^{|K_G|}}
\end{equation}
If the value of the trace distance is very large in comparison with 
$\frac{1}{2^{|K_G|}}$, the guessing probability is very large. 
So such a one time pad may be decrypted easily[5,6]. 

In addition, QKD needs an initial secret key for the authentification 
  before the legitimate users start the QKD protocol. 
This is the same situation as the conventional symmetric cipher in which   
the key is for initial seed key for PRNG. Thus, we cannot start cryptographic
 action without certain initial secret key, except for the conventional public 
 key encryption.
\\

\section {Definition and security of quantum enigma cipher}
\subsection{Definition}
Let us describe here the ideal quantum enigma cipher system.
The quantum enigma cipher consists of an integration of mathematical 
encryption box and physical randomization for ciphertext of mathematical 
encryption box [2].
The mathematical encryption box has a secret key of the length $|K_s|$ bits 
and PRNG for expansion of the secret key. The physical encryption box has 
a mechanism to create ciphertext as signal and it has a function to 
induces an error when the eavesdropper receives the ciphertext as signal.
Consequently  different ciphertext sequences are observed 
in the legitimate's
 receiver and the eavesdropper's receiver, respectively.
 A requirement for the physical randomization is 
 \begin{equation}
 P_e(Eve) >> P_e(Bob\quad or \quad Alice)
\end{equation}
This means that the error performance $P_e$ of the eavesdropper becomes
 worse than that of the legitimate user, when they observe the ciphertext
  as signal in  communication lines.
 We can consider many schemes to realize the above condition 
  based on several quantum effects such as quantum noise, entanglement. 
  But the system has to satisfy the conditions described 
  in the introduction.

\subsection{Security}
The conventional symmetric key cipher produces the ciphertext of length 
at most $2^{|K_s|}$ bits. Because the key length is $|K_s|$ bits, 
when the eavesdropper
 gets the known plaintext of the length $|K_s|$ bits and ciphertext 
 corresponding to them, she can pindown the secret key 
by the Brute force attack (trying $2^{|K_s|}$ key candidates).
That is, the guessing probability is one.
In addition, the sequence of the ciphertext has certain correlation because of 
the structure of PRNG. So the eavesdropper can investigate several 
mathematical algorithms to estimate the secret key.

In the ideal quantum enigma cipher, the eavesdropper's observation of 
the cipertext as signal in communication lines suffers error completely,
while the legitimate user does not. So the legitimate user can decrypt  
by the secret key, but the eavesdropper does not even if she gets the secret
 key after her observation of ciphertext as signal. 
 Thus, the guessing probability is\begin{equation}
P_G(K_s) = 2^{-|K_s|}
\end{equation}
even if she collects the ciphertext of $ 2^{|K_s|}$ bits. 
This means an immunity against the Brute force attack by computers.
On the other hand, the quantum no cloning theorem may protect 
a physical Brute force attack by cloning whole quantum states, 
because a set of quantum states for the quantum enigma cipher are designed by 
non-orthogonal state with very close signal distance each other. 
\\

\section{Application of quantum illumination}
It is not clear whether the quantum illumination protocol provides 
the ideal difference between the correctness of the ciphertext or not:
\begin{equation}
\eta_{ideal}=\frac{P_{G(Eve)}(K_s)}{P_{G(Bob)}(K_s)}
=P_{G(Eve)}(K_s)=2^{-|K_s|}
\end{equation}
However, it may be one of the quantum methods to create the different 
correctness of the ciphertext as signal according to [3,4]. If it is so, 
the application of quantum illumination is very easy.\\
\\
(1) Alice generates entangled state, and sends the signal mode to Bob.\\
(2) Bob prepares a mathematical encryption box to encrypt the data sequence
 and modulates the received light by means of BPSK, but the signal for the 
 modulator is the ciphertext from the mathematical encryption box.\\
(3) Bob employs the conventional optical amplifier to mask 
the ciphertext signal by the spontaneous emission noise from the amplifier.
Then he returns optical signal to Alice.\\
(4) Alice can recover the ciphertext signal from the masking by noise 
by means of entanglement effect at her receiver. But Eve's receiver
 suffers the error because she cannot use the entanglement.\\
 
So far, the quantum illumination protocol has been proposed 
for a direct encryption to plaintext, and for key generation [3,4]. 
However, this is not a good idea, because it is difficult to guarantee its
 security.
 It should be used as a physical randamoization technique. 
The scheme introduced here is an example based on the most simple
 cascade cipher of the mathematical encryption box and the physical 
 randomization.  Even so, the structure of security analysis is 
 drastically changed. First we need an optimaization as follows:
\begin{equation}
\eta_{error} =\max\limits_{\Lambda_{QI}} \frac{P_e(Eve)}{P_e(Alice)}
\end{equation}
where $\Lambda_{QI}$ is a set of quantum illumination with several physical 
parameters. According to [4], 
\begin{eqnarray}
P_e(Eve)&=&exp(-4W{\kappa_s}G_BN^2_S/RN_B)/2 \\
P_e(Alice)&=&exp(-W{\kappa_s^3 }G_BN_S/RN_B)/2
\end{eqnarray}
Although we need complicated analysis to derive the guessing probabiliy
 of the secret key of the mathematical enryption box based on the above 
 under the unconditional setting, 
 it may be expected to perform well.
 
 \section{For future}
 We have given an example of the concept of the quantum enigma cipher by using 
 quantum illumination technique.
In the sense of theoretical cryptology, such a cascade cipher 
 would not be attractive.
 The real quantum enigma cipher requires the emergence of the additional 
 enhancement for the security or the function by integrating the mathematical 
 encryption box and the physical randomization such as the triple DES 
 in the conventional cipher. The quantum noise randomized stream cipher 
 $\alpha/\eta$[8,9] and Y-00 [10,11] is a random  cipher along with this 
 concept. 
 However, the masking is very small in the real setting, so they need 
 additional randomization methods [12] or a new technique.
 A key concept is how to protect a secret key of the mathematical 
 encryption box by the law of quantum mechanics. The author is expecting  
 a good proposal.

\section{Conclusion}
Quantum key distribution has only just one of the functions of 
cryptology that provide secret key sequence 
for realizing one time pad. As described in this note, the one time pad 
is not a new concept in cryptology, but quantum enigma cipher is indeed 
a new concept in cryptology. It was claimed that the quantum illumination may 
provide G bit/sec rate. But the direct encryption and also 
the one time pad based on it are not an attractive scheme for the real world.
Thus it is preferable to adopt as a physical randomization technique 
to the quantum enigma cipher.

The purpose of this note is to introduce the concept of the quantum 
enigma cipher using the quantum illumination. 
The author would like to emphasize that 
there are many ways to realize quantum enigma cipher by 
applying the quantum physics, but he does not know how to establish 
the ideal system design to provide the different performance from 
simple cascade effect.
This note may give a hint.

\section*{Acknowledgment}
I am grateful to M.Sohma (Chief Professor of Quantum ICT Research Institute) 
 for fruitful discussions.



\end{document}